\documentclass[reprint, aps, prl, floatfix,superscriptaddress]{revtex4-2}

\usepackage{graphicx}    
\usepackage{dcolumn}    
\usepackage{bm}             
\usepackage{amssymb}   
\usepackage{amsmath}    
\usepackage{amsthm}
\usepackage{mathrsfs}    
\usepackage{commath}   
\usepackage{subfigure}    
\usepackage{braket}
\usepackage{float}
\usepackage{color}

\usepackage[colorlinks,
linkcolor=blue,
anchorcolor=blue,
citecolor=blue,
filecolor=blue,
menucolor=blue,
runcolor=blue,
urlcolor=blue,
frenchlinks=blue]{hyperref}

\usepackage[normalem]{ulem}
\usepackage{physics}

\begin{document}


\title{Probing topological phase transition with non-Hermitian  perturbations}
\author{Jingcheng Liang}
\email{liangjc@iphy.ac.cn}
\affiliation{Beijing National Laboratory for Condensed Matter Physics and Institute of Physics, Chinese Academy of Sciences, Beijing 100190, China}

\author{Chen Fang}
\email{cfang@iphy.ac.cn}
\affiliation{Beijing National Laboratory for Condensed Matter Physics and Institute of Physics, Chinese Academy of Sciences, Beijing 100190, China}


\author{Jiangping Hu}
\email{jphu@iphy.ac.cn}
\affiliation{Beijing National Laboratory for Condensed Matter Physics and Institute of Physics, Chinese Academy of Sciences, Beijing 100190, China}
\affiliation{Kavli Institute of Theoretical Sciences, University of Chinese Academy of Sciences,
	Beijing, 100190, China}
\affiliation{New Cornerstone Science Laboratory, Shenzhen 518054, China}




\date{\today}

\begin{abstract}
We demonstrate that non-Hermitian perturbations can probe topological phase transitions and unambiguously detect non-Abelian zero modes.  We show that under carefully designed non-Hermitian perturbations, the Loschmidt echo(LE) decays into 1/N where N is the ground state degeneracy in the topological non-trivial phase, while it approaches 1 in the trivial phase. This distinction is robust against small parameter deviations in the non-Hermitian perturbations. We further study four well-known models that support Majorana or parafermionic zero modes. By calculating their dynamical responses to specific non-Hermitian perturbations, we prove that the steady-state LE can indeed differentiate between different phases. This method avoids the ambiguity introduced by trivial zero-energy states and thus provides an alternative and promising way to demonstrate the emergence of topologically non-trivial phases. The experimental realizations of non-Hermitian perturbations are discussed.

\end{abstract}

\maketitle


\textit{Introduction}.---
Majorana zero modes (MZM), and the more exotic parafermion zero modes (PZM) are proposed to emerge in various systems and may be used as the building blocks of topological quantum computers\cite{Moore1991-rf,Read2000-lz,Yu_Kitaev2001-ys,Fu2008-yo,Lutchyn2010-xj,Oreg2010-qv,Sau2010-wi,Mong2014-lm,Nayak2008-nf,Alicea2012-vx,Clarke2013-ym}. These non-Abelian anyons and the related topological phase transitions are studied in a lot of experimental setups\cite{Wu2018-po,Rokhinson2012-js,Albrecht2016-ei,Fornieri2019-al,Mourik2012-hd,Pientka2017-kr,Das2012-hv,Wang2021-bo}. Previous experimental efforts mainly focus on the electromagnetic response signals, i.e., the zero-bias conductance peaks and the fractional Josephson effect\cite{Mourik2012-hd,Rokhinson2012-js}. However, there are suggestions that these experimental signatures are not sufficient to demonstrate the exsistence of Majorana zero modes, since they may be explained by other factors such as disorders, random fluctuation of superconducting gap, fine-tuning of experimental parameters and nearly zero-energy Andreev bound state in the trivial superconductivity regime\cite{Liu2012-fu,Liu2017-xm,ji20181,Pan2020-ph,Pan2020-uv,Chen2019-wi,Liu2021-ay}. Therefore, it is important to look for unique signatures of the MZM and PZM and the accompanying topological phase transition.

On the other hand, the non-Hermitian physics attracts much attention in recent years\cite{Bender1998-pu, Bender2007-qr, El-Ganainy2018-vd, Shen2018-pi, Kunst2018-ci, Yao2018-ob,Yao2018-lq, Longhi2019-rv, Kawabata2019-zp, Yokomizo2019-xo, Kawabata2019-ve, Okuma2019-tx,Kawabata2019-fc, Xiao2020-zb, Pan2020-zn, Ashida2020-ez, Zhang2020-yx, Yang2020-ap,Bergholtz2021-ng, Zhang2021-rb, Xiao2021-li, Zhang2021-if, Guo2021-ze, Sticlet2022-td}. These non-Hermitian systems are essentially open quantum systems where the influence of the environment is approximated by non-Hermitian terms in the Hamiltonian\cite{breuer2002theory, Datta2005-xt,Bender2007-qr, El-Ganainy2018-vd,Ashida2020-ez,Bergholtz2021-ng}. Introducing non-Hermiticity into topological systems lead to a plethora of novel phenomena such as non-Hermitian skin effect, breakdown of usual bulk boundary correspondence and sensitivity to boundary conditions\cite{Guo2021-ze, Yang2020-ap, Kunst2018-ci,Okuma2019-tx, Yokomizo2019-xo, Yao2018-ob}. Moreover, the linear response theory of non-Hermitian probes is studied in Refs.~\cite{Sticlet2022-td, Pan2020-zn} and peculiar properties such slower dynamics and tachyon phase are found. Recently, the dynamical response of non-Hermitian perturbations in different quantum phases are studies in Refs.~\cite{Zhang2021-if, Zhang2021-rb}. They find that the Loschmidt echo(LE) which characterize the dynamical effect of the perturbation can be used to differentiate different quantum phases. This distinction is due to the peculiar dynamics behavior of a degenerate subspace near the exceptional points(EP). Hence, it is natural to ask whether this approach can be generalized to the regime of topological phase transitions where there is also a state degeneracy. If such a generalization is possible, it will bring about significant advancements to the field of topological quantum computations by providing an alternative and promising way to unambiguously demonstrate the existence of MZM and PZM.

In this letter, we show that the steady state values of Loschmidt echo(LE) indeed can be used to differentiate topological trivial and non-trivial phases. While Refs.~\cite{Zhang2021-if, Zhang2021-rb} only prove the effectiveness of this method for doubly degenerate systems, we demonstrate its validity for arbitrary degenerate systems. We further investigate the effects of random perturbations and find that this method is quite robust against small deviations, which renders flexibility when choosing the suitable non-Hermitian probe. We then study four models that can host MZM or PZM: the Kitaev chain, coupled Kitaev chains, a semiconductor nanowire in proximity to s-wave superconductor and a parafermion model based on the interplay of superconductivity and fractional quantum Hall effect. After introducing a carefully designed non-Hermitian perturbation, which assumes the Jordan bock form in the degenerate subspace, into the post-quenched Hamiltonian, the steady state LE approaches $\frac{1}{N}$ in the topological non-trivial phase where N is the ground state degeneracy in the specific models, while the LE fluctuates around 1 in the topological trivial phase. This feature remains true even if this specific non-Hemitian perturbation does not make the post-quench Hamiltonian singular. Therefore, a decrease in the steady state LE values can be used as an indication of the topological non-trivial phase. The specific non-Hermitian perturbation may arise as a part of the effective Hamiltonian that describes the quasiparticle poisoning process or the non-Markovian tunneling, which change the fermion parity of the subsystems while conserve the parity of the total system\cite{Colbert2014-ux, Budich2012-gk, Goldstein2011-np, Rainis2012-yc, Karzig2021-bo,Breuer2016-zh}. We emphasize that the decrease in LE under such specific non-Hermitian perturbations depends on the overall energy structure of the Hamiltonian and is not affected by trivial zero energy states. Therefore, our work offers a promising alternative method for distinguishing MZM and PZM from trivial zero energy states and for unambiguously demonstrating topological phase transitions.

\textit{Dynamics under non-Hermitian perturbations}.---
The main quantity we explore in this paper is the Loschmidt echo (or Uhlmann fidelity)\cite{Zhang2021-rb},
\begin{eqnarray}
L(t)=(\Tr\sqrt{\sqrt{\rho(0)}\rho(t)\sqrt{\rho(0)}})^2
\label{Lt}
\end{eqnarray}
where $\rho(0)$ is the initial density matrix and $\rho(t)$ is the time evolved density matrix given by 
\begin{eqnarray}
\rho(t)=e^{-i\mathcal{H}t}\rho(0)e^{i\mathcal{H}^\dagger t}
\label{rho_t}
\end{eqnarray}
The post-quench Hamiltonian $\mathcal{H}$ 
\begin{eqnarray}
    \mathcal{H}=H+\lambda J
\end{eqnarray}
where $H$ is the Hamiltonian that host MZM or PZM, $J$ is the non-Hermitian perturbation term. 

For the topological non-trivial phase, all the energy levels are N-fold degenerate, where $N=2$ corresponds to the MZM. Within the degenerate subspace, $[H]=EI_N$ and we take $[J]_{ij}=\delta_{i-1,j}$, then $\mathcal{H}$ is at the exceptional point. We denote as $\ket{\psi_{in}}$ the column vector that has only one nonzero value 1 on its nth entry. It is readily to verify that $\ket{\psi_{iN}}$ is the only eigenstate of $\mathcal{H}$. For an arbitrary initial state in this subspace $\ket{\psi(0)}=\sum_{m=1}^Na_m\ket{\psi_{im}}$, the time evolved state is given by\cite{supp} 
\begin{eqnarray}
    \ket{\psi(t)}=e^{-iEt}\sum_{m=1}^N(\sum_{n=0}^{m-1}\frac{(-i\lambda t)^n}{n!}a_{m-n})\ket{\psi_{im}}
\end{eqnarray}
Therefore, for $t\gg 1/\lambda$, any initial state will evolve to $\ket{\psi_{iN}}$ leading to $\rho(t)=\ketbra{\psi_{iN}}$. If the system is initially prepared in the thermal equilibrium state, then we have $\rho(0)=I_N/N$. Using Eq.~(\ref{Lt}), we find steady state $L(t)=1/N$.

For the topological trivial phase, if we assume that the inter-level couplings are small, we may still only need to consider the dynamics of the same subspace. The only change is now $[H]=\text{diag}\{E_1, E_2, ..., E_N\}$. For $\rho(0)=I_N/N$, we have\cite{supp}
\begin{eqnarray}
    \rho(t)\approx \frac{1}{N}(I_N-i\lambda\int_0^t J_I(t')dt'+i\lambda\int_0^t J_I^\dagger(t')dt')
    \label{rho_t1}
\end{eqnarray}
where $J_I(t')=e^{-iHt'}Je^{iHt'}$. The integration in Eq.~(\ref{rho_t1}) gives a factor $1/\Delta E$ where $\Delta E$ is the splitting of energy of $H$. Thus, we can estimate that the steady state $L(t)\approx 1+O(\lambda/\Delta E)$. As long as the non-Hermitian perturbation is small compared to the splitting of energy of $H$, $L(t)$ is near 1, which is drastically different from its value $1/N$ in the topological non-trivial phase. Therefore, the steady state LE value can be used to probe different phases.

Now, let's consider the robustness of this distinction. The post-quench Hamiltonian now becomes
\begin{eqnarray}
    \mathcal{H}=H+\lambda J+\lambda' J^\dagger
\end{eqnarray}
where $[J]_{ij}=\delta_{i-1,j}$ in the degenerate subspace. The reason why we choose $J^\dagger$ as the random perturbation is that it makes $\mathcal{H}$ continuously connected to the Hermitian case and it represents the most common type of random perturbations. For MZM supporting systems where the dimension of degenerate subspace is 2, the $J^\dagger$ term is the only possible random fermion-parity changing perturbation. For a given $\rho(0)$, we can analytically obtain $\rho(t)$ from Eq.~(\ref{rho_t}).
 We find that for small random perturbation $\lambda'\ll \lambda$, if $1/\sqrt{\lambda\lambda'}\gg t\gg 1/\lambda$, then $L(t)\approx 1/2$ in topological non-trivial phase. See Supplementary Materials for detailed derivations. 

For topological non-trivial phase with general degenerate subspace dimension N, we have $\rho(t)=\frac{1}{N}e^{-it(\lambda J+\lambda' J^\dagger)}e^{it(\lambda' J+\lambda J^\dagger)}$. When N is large, we may approximate $[J, J^\dagger]\approx 0$. Then, for $1/\lambda'\gg t\gg 1/\lambda$ and remembering $J$ and $J^\dagger$ are all nilpotent matrix with index N, we see that $\rho_{NN}$ is the dominant component of $\rho(t)$ which leads to $L(t)\approx 1/N$ according to Eq.~(\ref{Lt}).

For topological trivial phase with energy splitting $\Delta E\gg\lambda, \lambda'$, we can still apply the perturbation expansion as Eq.~(\ref{rho_t1}). The deviation of $\rho(t)$ from $I_N/N$ is proportional to $(\lambda-\lambda')/\Delta E$ (see \cite{supp}). Therefore, for small random perturbations, we always have steady state $L(t)\approx1+O(\lambda/\Delta E)\approx 1$. 

The robustness of different behaviors of $L(t)$ in different phases provides us flexibility on choosing the non-Hermitian probes for practical models. We will see in the numerical simulations for semiconductor nanowire that the distinction is preserved for quite general non-Hermitian perturbations\cite{supp}. This feature makes this method easier to realize experimentally since fine-tuning of Hamiltonian is not necessary.

\textit{Probing the topological phase transition in MZM models}.---To further verify our method of using steady state Loschmidt echo to probe topological phase transition, we conduct numerical simulations on several practical models. We consider first the celebrated Kitaev chain model for one dimensional p-wave superconductivity\cite{Yu_Kitaev2001-ys}. The Hamiltonian is given by
\begin{eqnarray}
H=\sum_j^{N-1} [&-&t(a_j^\dagger a_{j+1}+a_{j+1}^\dagger a_j)-\mu(a_j^\dagger a_j-1/2)\nonumber\\
&+&\Delta a_j a_{j+1}
+\Delta^*a_{j+1}^\dagger a_j^\dagger]
\label{Kitaev}
\end{eqnarray}
where t is the hopping amplitude, $\mu$ the chemical potential and $\Delta$ the superconducting pairing potential which is assumed to be real and positive. 

\begin{figure}[!t]
    \centering
    \includegraphics[width=3.35in]{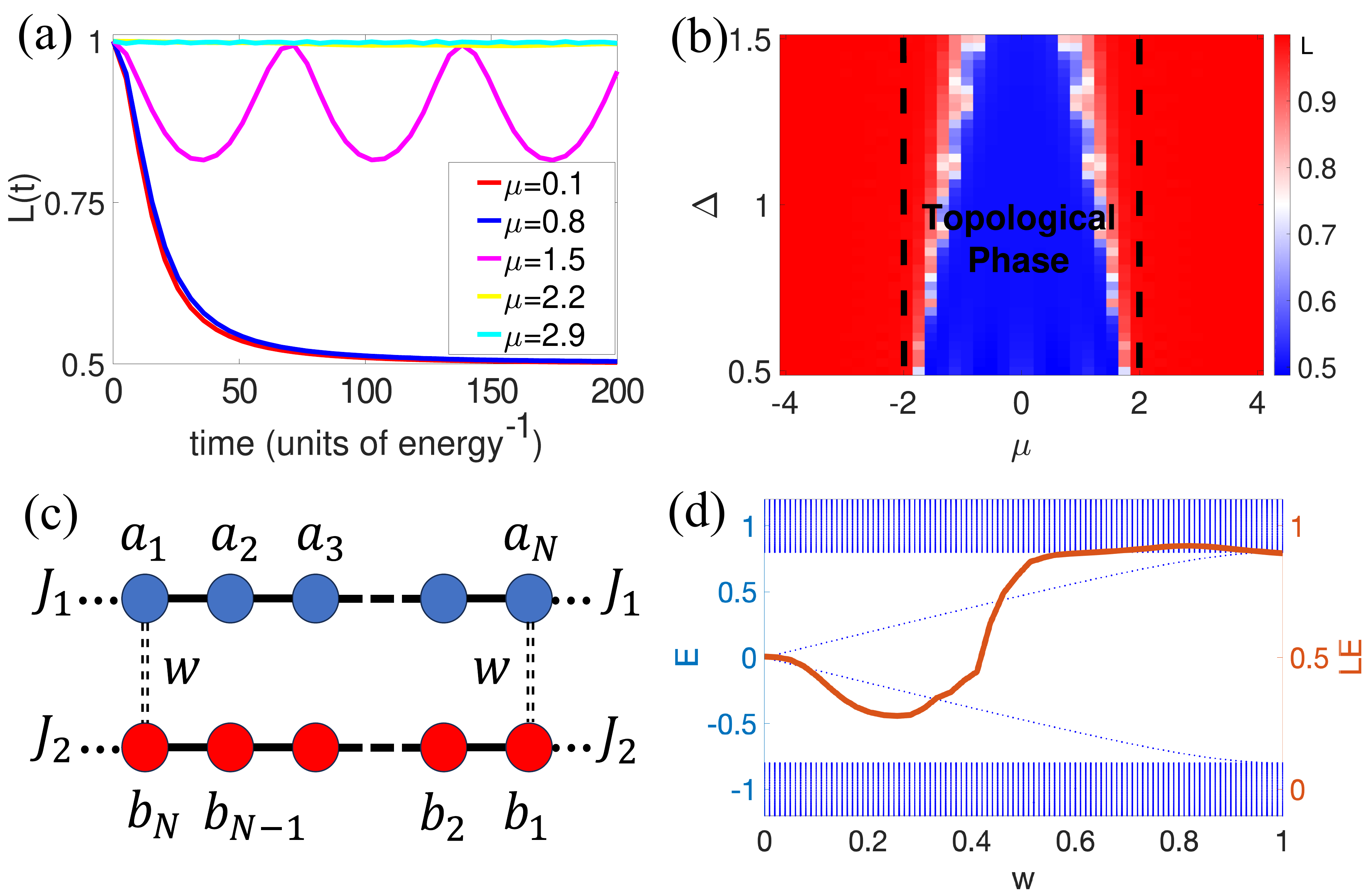}
    
    \caption{(a)(b) are numerical simulations for Kitaev chain model and (c)(d) are for double Kitaev chains. (a) The time evolution of  $L(t)$ for different $\mu$ values. Parameters: $t=1$, $\Delta=1$, $\beta=5$, $\lambda=0.1$. The total number of sites of the chain is $N=8$. The phase boundary is $\mu=2t=2$. we observe that the steady state $L(t)$ indeed converge to 1 and 0.5 for topological trivial and non-trivial phases, respectively. (b) The phase diagram obtained from $\bar{L}$ for different values of $\Delta$ and $\mu$. The dash lines represent the theoretical phase boundary $|\mu|=2|t|$. The other parameters are the same as (a). (c)A schematic illustration of the double Kitaev chains. (d)The energy spectrum(blue dots) and steady state LE value $\bar{L}$ (red curve) for varying coupling strength $w$ between the two chains. Parameters: $t=1$, $\Delta=1$, $\beta=5$, $\mu=0.4$, $\kappa_1=\kappa_2=0.1$. }
    \label{fig1}
\end{figure}

 Now we need to find out the suitable non-Hermitian term that takes the form $[J]_{ij}=\delta_{i-1,j}$ in the degenerate subspace. $J$ should change fermionic parity and satisfies the commutation relation
\begin{eqnarray}
[J, P_\gamma]=\pm 2 J
\label{condition1}
\end{eqnarray}
where $P_\gamma$ is the fermionic parity operator defined in \cite{supp}. For $\mu\approx0$ and $t\approx\Delta$, the MZM $\gamma$ and $\gamma'$ are approximated by $\gamma\approx c_1$ and $\gamma'\approx c_{2N}$ which are well localized on the edges of the chain\cite{supp}. Substuting $P_\gamma=i c_1 c_{2N}$ into Eq.~(\ref{condition1}), we have
\begin{eqnarray}
J=a_1+a_1^\dagger\pm (a_N-a_N^\dagger)
\label{NH1}
\end{eqnarray}

We choose a particular branch $J=a_1+a_1^\dagger+a_N-a_N^\dagger$ to simulate the dynamics. We assume the initial state is a thermal state with density matrix $\rho(0)=e^{-\beta H}/\Tr(e^{-\beta H})$ and it evolves under the quenched Hamiltonian $\mathcal{H}=H+\lambda J$. The normalized density matrix $\rho(t)$ at time t is given by\cite{Kawabata2017-mm, Brody2012-rf, Zhang2021-rb}
\begin{eqnarray}
\rho(t)=e^{-i\mathcal{H}t}\rho(0)e^{i\mathcal{H}^\dagger t}/\Tr(e^{-i\mathcal{H}t}\rho(0)e^{i\mathcal{H}^\dagger t})
\label{rho_t2}
\end{eqnarray}
$L(t)$ is obtained using Eq.~(\ref{Lt}). Different dynamical behaviors of $L(t)$ mark different phases. Varying the chemical potential $\mu$, the numerical results of $L(t)$ is given in Fig.~\ref{fig1}(a). For $\mu=0.1$ and $\mu=0.8$ which are well resided in the topological non-trivial regime, the $L(t)$ indeed tend to 0.5 as expected. For $\mu=1.5$ which is close to the phase boundary, the $L(t)$ fluctuates around 0.9. For topological trivial phase with $\mu=2.2$ and $\mu=2.9$, the $L(t)$ remains nearly 1. Therefore, the steady state LE indeed can be used to differentiate different phases.

To explore a broader range of parameters, we define an average steady state $\bar{L}$\cite{Zhang2021-rb}
\begin{eqnarray}
\bar{L}=\frac{1}{t_1-t_0}\int_{t_0}^{t_1}L(t')dt'
\label{L_bar}
\end{eqnarray}
Vary both the $\Delta$ and $\mu$, the numerical results of $\bar{L}$ are plotted in Fig.~\ref{fig1}(b). The dash lines mark the theoretical phase boundary. We see that $\bar{L}$ approximately predicts the right phase with errors around the phase boundary. These errors can be traced back to our choice of an approximate $H_\delta$ in Eq.~(\ref{NH1}) for specific parameters. A detail discussion for the error is given in \cite{supp}. From Fig.~\ref{fig1}(b), we conclude that, except for the points near phase boundary, steady state $L(t)$ can be used to detect the topological phase in a wide range of parameter space and provides an effective method to explore the Majorana zero modes.

To further explore the relationship between $\bar{L}$ and MZM, we consider the double Kitaev chain model schematically shown in Fig.~\ref{fig1}(c). The annihilation operator of the two chains are represented by $a$ and $b$ respectively. The Hamiltonian of each chain is the same as Eq.~(\ref{Kitaev}) with tunnellings:
\begin{align}
    \delta H&=-w(a_N^\dagger b_1+b_1^\dagger a_N+b^\dagger_N a_1+a_1^\dagger b_N\nonumber\\
    &-a_Nb_1-b_1^\dagger a_N^\dagger-b_Na_1-a^\dagger_1b_N^\dagger)
\end{align}
The non-Hermitian perturbation is given by $J_1=\kappa_1 (a_1+a_1^\dagger+a_N-a_N^\dagger)$ and $J_2=\kappa_2 (b_1+b_1^\dagger+b_N-b_N^\dagger)$. When $w$ is tuned from 0 to 1, from Fig.~\ref{fig1}(d) we find that the gap between zero modes becomes larger and larger, and eventually merge into the bulk states at $w=1$. Meanwhile, the $\bar{L}$(calculated in the range $t_0=100$, $t_1=200$) starts from 0.5, then decreases for small $w$ and approaches 1 for large $w$. This result is consistent with our expectation that a large value of $\bar{L}\approx 1$ is a NULL signal for MZM. The initial decreasing of $\bar{L}$ may be due to the fact that when the gap opening is not too large, the ground state degeneracy of the two chains is nearly 4. But our $J_1$ and $J_2$ do not take the Jordan block form in the new nearly-degenerate subspace, the $\bar{L}$ just decreases from 0.5 and does not approach 1/4.

Now we consider a more realistic model to realize MZM which utilizes the semiconductor-superconductor hetero-structures\cite{Lutchyn2010-xj,Oreg2010-qv,Sau2010-wi}. We consider a one-dimensional semiconductor wire with strong Rashba spin orbit interaction placed in proximity to a superconductor. 
From the continuum Hamiltonian, we obtain the tight binding Hamiltonian:
\begin{eqnarray}
H&=&\sum_{i,\sigma}\frac{1}{2}[(2t-\mu)a_{i\sigma}^\dagger a_{i\sigma}-(2t-2i\alpha\sigma)a_{i+1\sigma}^\dagger a_{i\sigma}\nonumber\\
&+&V a_{i\uparrow}^\dagger a_{i\downarrow}+\Delta a_{i\uparrow} a_{i\downarrow}]+H.c.
\label{Hi_spin_main}
\end{eqnarray}
where $\sigma$ represents the spin. See Supplementary Materials for detailed derivation.

\begin{figure}[!t]
    \centering
    \includegraphics[width=3.35in]{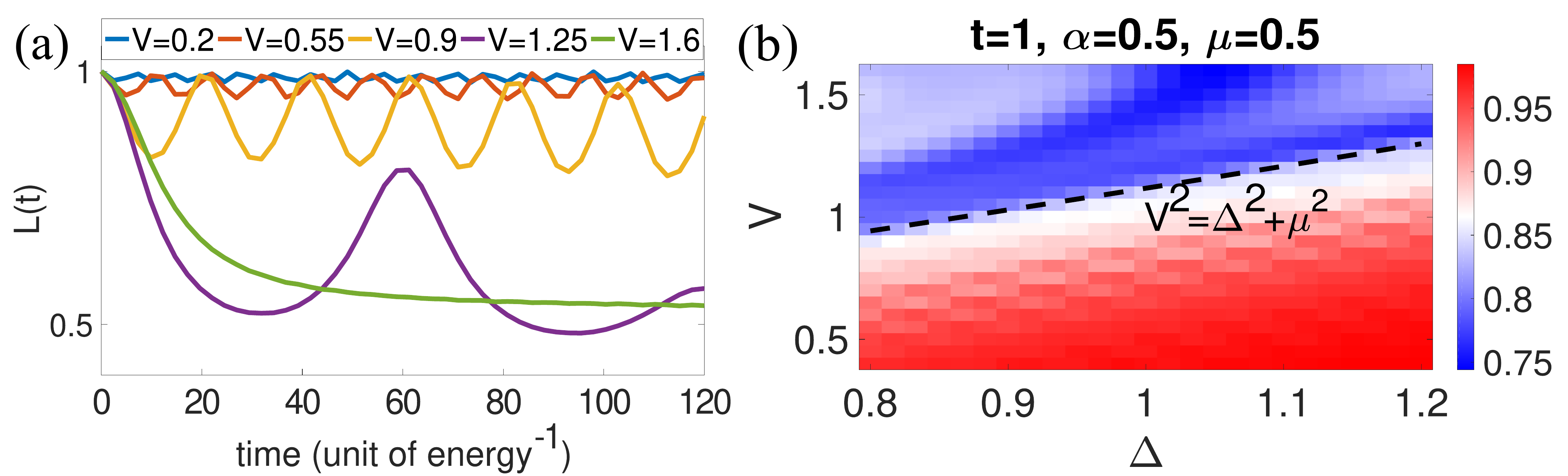}
    
    \caption{(a) The time evolution of $L(t)$ for different V values. Parameters: $t=1$, $\Delta=1$, $\mu=0.5$, $\alpha=0.5$, $\beta=5$, $\lambda=0.05$. The total number of lattice sites is $N=5$. The topological phase transition happens at $V_c=\sqrt{\Delta^2+\mu^2}\approx 1.12$. For topological trivial phase $V<V_c$, $L(t)$ fluctuates around 1. For topological non-trivial phase $V>V_c$, $L(t)$ evolves to a steady state value close to 0.5. The larger fluctuations of $V=1.25$ case may be due to the size effect in numerical simulation. (b)Phase diagram obtained from the steady state value $\bar{L}$ with non-Hermitian perturbation given by $J=\gamma+i\gamma'$ according to Eq.~(\ref{MZM_1}). Parameters: $t=1$, $\mu=0.5$, $\alpha=0.5$, $\beta=5$, $N=4$.The dash line represents the theoretical phase boundary.}
    \label{fig2}
\end{figure}

Now, to differentiate different phases through the LE, we need to find out the non-Hermitian perturbation $J$ that satisfies Eq.~(\ref{condition1}) as before. However, in the current model, we need to solve Eq.~(\ref{Hi_spin_main}) numerically to obtain MZM for a specific set of parameters. 
We choose parameters: $t=1$, $\Delta=1$, $\mu=0.5$, $\alpha=0.5$, $V=1.5$, substitute them into Eq.~(\ref{Hi_spin_main}) with total number of lattice sites $N=50$ and keep only the edge components, we obtain,
\begin{eqnarray}
\gamma&\approx&ie^{i\phi}a_{1\uparrow}^\dagger-ie^{-i\phi}a_{1\uparrow}+e^{-i\phi}a_{1\downarrow}^\dagger+e^{i\phi}a_{1\downarrow}\nonumber\\
\gamma'&\approx&ie^{-i\phi}a_{N\uparrow}^\dagger-ie^{i\phi}a_{N\uparrow}-e^{i\phi}a_{N\downarrow}^\dagger-e^{-i\phi}a_{N\downarrow}
\label{MZM_1}
\end{eqnarray}
where $\phi=0.36$. The $J$ that satisfies Eq.~(\ref{condition1}) is $\gamma\pm i\gamma'$. We choose
$J=\gamma+i\gamma'$. The post-quenched Hamiltonian $\mathcal{H}=H+\lambda J$. Gradually increasing the Zeeman splitting value from $V=0.2$ to $V=1.6$ , using Eq.~(\ref{Lt}) and Eq.~(\ref{rho_t}), we obtain the corresponding $L(t)$ given in Fig.~\ref{fig2}(a). It can be seen from Fig.~\ref{fig2}(a) that the steady state $L(t)$ again take values as expected in different phases. Our numerical simulations only involves $N=5$ lattice sites, therefore the Majorana zero modes have large overlap, leading to large fluctuations.

To see if $J$ can be used to explore the phases of larger parameter space, we vary both $V$ and $\Delta$ and calculate the corresponding $\bar{L}$ using Eq.~(\ref{L_bar}) with $t_0=90$ and $t_1=110$. The results are given in Fig.~\ref{fig2}(b). Although we only have lattice sites $N=4$, The phase diagram still predicts the correct phase and the phase boundary. 

\begin{figure*}
    \centering
    \includegraphics[width=7in]{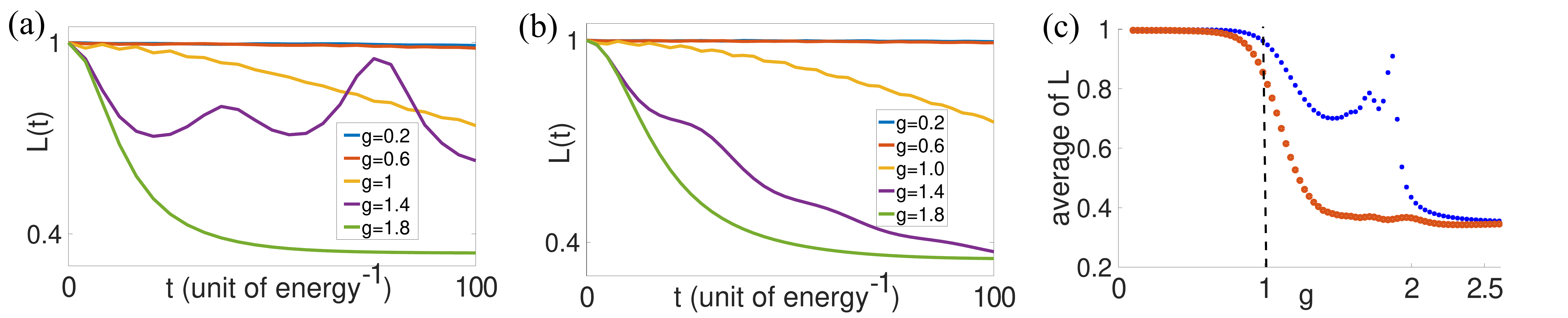}
    \caption{(a)Evolution of $L(t)$ under $\mathcal{H}_1$ for different values of g in Eq.~(\ref{para_H}). The steady state $L(t)$ indeed approaches $\frac{1}{3}$ in the topological phase. The large fluctuations of the $g=1.4$ case may be due the approximation in the $H_\delta$ and the finite size of the parafermion chain. Other parameters: $h=1$, $\beta=5$, $N=6$, $\lambda=0.05$.(b)$L(t)$ under $\mathcal{H}_2$ for different values of g in Eq.~(\ref{para_H}). The $g=1.4$ is now smooth and approaches $\frac{1}{3}$. (c)The steady state value $\bar{L}$ as a function of g. The blue symbols represent the results obtained under $\mathcal{H}_1$, while the red symbols represent the results obtained under $\mathcal{H}_2$. As expected, the $\bar{L}$ starts to decrease around the critical point $g=1$ and for large g, $\bar{L}$ approaches $\frac{1}{3}$. The $\mathcal{H}_2$ case with an ideal Jordan block form predicts a better curve for the phase transition. Other parameters: $h=1$, $\beta=5$, $N=5$, $\lambda=0.05$.}
    \label{fig3}
\end{figure*}

The above method requires solving the exact Majorana zero modes Eq.~(\ref{MZM_1}) which may be impractical in experiments. However, we may still use this dynamical method to differentiate the topological trivial and non-trivial phases by choosing a non-Hermitian perturbation that contains contributions from both edges. The derivation and numerical results of this point are given in the Supplementary Materials. Therefore it renders more flexibility to the experimental realizations.

\textit{Probing the topological phase transitions in parafermion models}.---It is well known that Majorana zero modes can not approximate all unitary quantum gates\cite{Freedman2002-xb,Baraban2010-th}. To overcome this problem, the parafermion zero modes have been proposed as building blocks for realizations of universal braiding operations\cite{Mong2014-lm,Nayak2008-nf,Read1999-wq,Fendley2012-rr,Alicea2016-zt}. The parafermion zero mode is a generalization of the Majorana zero mode which has a more exotic algebra structure:
$\alpha_j^d=1$, $\alpha_j^\dagger=\alpha_j^{d-1}$ and 
$\alpha_j\alpha_{j'}=\alpha_{j'}\alpha_je^{i(2\pi/d)sgn(j'-j)}$. We only explore the $d=3$ case in this letter.


We consider a general one dimensional parafermion system whose Hamiltonian is 
\begin{eqnarray}
H=&h&\sum_{j=1}^N(e^{i\frac{\pi}{d}}\alpha_{2j-1}^\dagger\alpha_{2j}+H.c.)\nonumber\\
&+&g\sum_{j=1}^{N-1}(e^{-i\frac{\pi}{d}}\alpha_{2j}^\dagger\alpha_{2j+1}+H.c.)
\label{para_H}
\end{eqnarray}

Now we are to find out the non-Hermitian perturbation $J$ which takes the Jordan block form $[J]_{i,j}=\delta_{i-1,j}$ within the ground state subspace. Approximately we may choose $J_1=\alpha_1$. 
We may also choose one of the operators that take the exact Jordan block form $J_0=\alpha_1+e^{-i\pi/3}\alpha_{2N}$. In our numerical simulations, we explore both the approximate one $\mathcal{H}_1=H+\lambda J_1$ and the exact one $\mathcal{H}_2=H+\lambda J_0$. In the Eq.~(\ref{para_H}), we let $h=1$, then $g=1$ is the phase boundary. Following the same procedure as previous sections, we obtain the time evolution of LE using $\mathcal{H}_1$ for different values of g in Fig.~\ref{fig3}(a). Because the ground state degeneracy in the ordered phase is now $d=3$, we expect the $\bar{L}$ approaches $\frac{1}{3}$. This is exactly the case of $g=1.8$. The $g=1.4$ case is also in the ordered phase but it does not approach $\frac{1}{3}$. The reason may be that $J_1$ is approximate and the size of the chain is only $N=6$. The results of $\mathcal{H}_2$ are given in Fig.~\ref{fig3}(b). We observe that the case $g=1.4$ is now smooth and the steady value approaches $\frac{1}{3}$. We also plot the steady state value $\bar{L}$ for different g of both $\mathcal{H}_1$ and $\mathcal{H}_2$ in Fig.~\ref{fig3}(c). It also indicates that a phase transition will happen by varying g in Eq.~(\ref{para_H}). Therefore, our dynamical method for probing phase transitions remains valid for this parafermion model. It may serve as an alternative way to study the emergence of parafermion zero modes in various systems.

\textit{Discussions and conclusions}.---
The realization of the required Jordan block like non-Hermitian Hamiltonian may be a challenge in experiments. However, if the MZM or PZM supporting systems are open systems, we learn that the quasiparticle poisoning process or the non-Markovian tunneling can change the fermion parity of the subsystems while conserve the parity of the total system\cite{Colbert2014-ux, Budich2012-gk, Goldstein2011-np, Rainis2012-yc, Karzig2021-bo,Breuer2016-zh}.We anticipate our desired non-Hermitian Hamiltonian may arise as an effective Hamiltonian in the quantum master equations that describe the dynamics of these open systems. 

For the measurement of the LE of a many-body system, Ref.~\cite{Vasseur2014-mo} mentions several proposals, including x-ray edge setup\cite{mahan1967excitons,nozieres1969singularities}, quantum dot optical absorption\cite{tureci2011many,vasseur2014edge} and Ramsey interferometry techniques\cite{knap2012time,dora2013loschmidt}. 

In conclusion, through the studies of different models of MZM and PZM, we have seen that a suitable choice of non-Hermitian perturbations indeed leads to different dynamical behaviors in different phases. The steady state LE value can be used as an indicator for the corresponding phases as well as the presence of MZM and PZM. This approach not only provides an alternative way to probe the topological phase transitions and validate the emergence of Majorana and parafermion zero modes, it also avoids the problem related to the trivial zero energy states discussed in Ref.~\cite{Liu2012-fu,Liu2017-xm,ji20181,Pan2020-ph,Pan2020-uv,Chen2019-wi,Liu2021-ay}. The reason is that, our dynamical approach actually probe the degeneracy in the whole energy spectrum. Trivial zero energy states only affect the ground state subspace but not the excited states, therefore they will not change the steady state values of $L(t)$. In this sense, our dynamical method is better than the conventional transport measurement approaches since it can unambiguously demonstrate the emergence of the non-trivial zero modes. 

\begin{acknowledgments}
The authors want to thank Kun Jiang, Tianyu Li, Yuncheng Xiong, Shu Chen, Haiping Hu, Zhesen Yang, Yuguo Liu for helpful discussions. This work is supported by the Ministry of Science and Technology  (Grant No. 2022YFA1403901), the National Natural Science Foundation of China (Grant No. NSFC-11888101), the Strategic Priority Research Program of the Chinese Academy of Sciences (Grant No. XDB28000000), the New Cornerstone Science Foundation, the International Young Scientist Fellowship of Institute of Physics Chinese Academy of Science (No. 202302).
\end{acknowledgments}

\bibliography{reference}
\newpage
\begin{widetext}

\appendix

\section{Appendix A: Derivation of the dynamics under non-Hermitian perturbations}
In this section, we provide a detailed derivation of the Eq.~(4) and Eq.~(5) in the main text. 

The Eq.~(4) comes from the fact that $J$ is a nilpotent matrix with index N,
\begin{eqnarray}
\ket{\psi(t)}&=&\mathcal{U}_i(t)\ket{\psi(0)}\nonumber\\
&=&e^{-i\mathcal{H}_it}\ket{\psi(0)}\nonumber\\
&=&e^{-iE_it}e^{-i\lambda J_it}\ket{\psi(0)}\nonumber\\
&=&e^{-iE_it}\sum_{n=0}^{N-1}\frac{(-i\lambda J_it)^n}{n!}\ket{\psi(0)}\nonumber\\
&=&e^{-iE_it}\sum_{m=1}^N(\sum_{n=0}^{m-1}\frac{(-i\lambda t)^n}{n!}a_{m-n})\ket{\psi_{im}}
\label{psi_t}
\end{eqnarray}

For the Eq.~(5) in the main text, we first note that $\rho(t)=\frac{1}{N}e^{-i\mathcal{H}t}e^{i\mathcal{H}^\dagger t}$. We have
\begin{eqnarray}
    e^{-i\mathcal{H}t}=\tilde{T}e^{-i\lambda\int_0^t J_I(t')dt'}e^{-iHt}
\end{eqnarray}
\begin{eqnarray}
    e^{i\mathcal{H}^\dagger t}=e^{iHt}Te^{i\lambda\int_0^t J_I^\dagger (t')dt'}
\end{eqnarray}
with $O(t')=e^{-iHt'}Oe^{iHt'}$, $T$ the time order operator and $\tilde{T}$ the anti-time order operator. Keeping terms up to first order leads to Eq.~(5). We further note that the non-zero terms in Eq.~(5) are 
\begin{eqnarray}
    &&-i\lambda \int_0^t[J_I(t')]_{i,i-1}dt'\nonumber\\
    &=&-i\lambda\int_0^t e^{-i(E_i-E_{i-1})t'}dt'\nonumber\\
    &=&\frac{\lambda}{E_i-E_{i-1}}(e^{-i(E_i-E_{i-1})t}-1)
\end{eqnarray}
and its Hermitian conjugate 
\begin{eqnarray}
    &&i\lambda \int_0^t[J_I^\dagger(t')]_{i-1,i}dt'\nonumber\\
    &=&\frac{\lambda}{E_i-E_{i-1}}(e^{i(E_i-E_{i-1})t}-1)
\end{eqnarray}
Therefore, the difference between $\rho(t)$ and $I_N/N$ is of order $\frac{\lambda}{\Delta E}$ which leads to an error of order $\frac{\lambda}{\Delta E}$ in steady state $L(t)$.

\section{Appendix B: The robustness of the distinction under random non-Hermitian perturbations}
In this section, we provide a detailed derivation for $\rho(t)$ in MZM case and deviation of $L(t)$ from 1 in topological trivial phase, while a random term is introduced in to the post-quench Hamiltonian $\mathcal{H}$:
\begin{eqnarray}
    \mathcal{H}=H+\lambda J+\lambda' J^\dagger
\end{eqnarray}
where $[J]_{ij}=\delta_{i-1,j}$ in the degenerate subspace. The reason why we choose $J^\dagger$ as the random perturbation is that it makes $\mathcal{H}$ continuously connected to the Hermitian case and it represents the most common type of random perturbations. For MZM supporting systems in their topological non-trivial phase, the dimension of degenerate subspace is $N=2$. In this case, the $J^\dagger$ term is the only possible random fermion-parity changing perturbation. Within the degenerate subspace, the Hamiltonian is 
\begin{eqnarray}
    \mathcal{H}=\left ( \begin{array}{cc}
         E&  \lambda'\\
        \lambda & E
    \end{array} \right)=EI_2+D
\end{eqnarray}
For $\rho(0)=I_2/2$, we have
\begin{eqnarray}
    \rho(t)&=&\frac{1}{2}e^{-i\mathcal{H}t}e^{i\mathcal{H}^\dagger t}\nonumber\\
    &=&\frac{1}{2}e^{-iDt}e^{iD^\dagger t}
    \label{app_rho_t11}
\end{eqnarray}
using $D^2=\lambda\lambda' I_2$, we have
\begin{eqnarray}
    e^{-iDt}&=&\cos(\sqrt{\lambda\lambda'}t)-\frac{iD}{\sqrt{\lambda\lambda'}}\sin(\sqrt{\lambda\lambda'}t)\nonumber\\
    e^{iD^\dagger t}&=&\cos(\sqrt{\lambda\lambda'}t)+\frac{iD^\dagger}{\sqrt{\lambda\lambda'}}\sin(\sqrt{\lambda\lambda'}t)
\end{eqnarray}
Substituting into Eq.~(\ref{app_rho_t11}), we have
\begin{eqnarray}
    \rho(t)=\frac{1}{2}\left ( \begin{array}{cc}
         \cos^2(\sqrt{\lambda\lambda'}t)+\frac{\lambda'}{\lambda}\sin^2(\sqrt{\lambda\lambda'}t)&  \frac{i}{2}(\sqrt{\frac{\lambda}{\lambda'}}-\sqrt{\frac{\lambda'}{\lambda}})\sin(2\sqrt{\lambda\lambda'}t)\\
        -\frac{i}{2}(\sqrt{\frac{\lambda}{\lambda'}}-\sqrt{\frac{\lambda'}{\lambda}})\sin(2\sqrt{\lambda\lambda'}t) & \cos^2(\sqrt{\lambda\lambda'}t)+\frac{\lambda'}{\lambda}\sin^2(\sqrt{\lambda\lambda'}t)
    \end{array} \right)
\end{eqnarray}
It is readily seen that for Hermitian case $\lambda'\approx \lambda$, we have $\rho(t)\approx I_2/2$ and $L(t)\approx 1$. For small random perturbation $\lambda'\ll \lambda$, if $1/\sqrt{\lambda\lambda'}\gg t\gg 1/\lambda$, then $\rho_{22}\gg\rho_{11}$ leading to $L(t)\approx 1/2$. 

For topological non-trivial phase with general degenerate subspace dimension N, we have $\rho(t)=\frac{1}{N}e^{-it(\lambda J+\lambda' J^\dagger)}e^{it(\lambda' J+\lambda J^\dagger)}$. When N is large, we may approximate $[J, J^\dagger]\approx 0$. Then, for $1/\lambda'\gg t\gg 1/\lambda$ and remembering $J$ and $J^\dagger$ are all nilpotent matrix with index N, we see that $\rho_{NN}$ is the dominant component of $\rho(t)$ which leads to $L(t)\approx 1/N$ according to Eq.~(\ref{Lt}).

For topological trivial phase with energy splitting $\Delta E\gg\lambda, \lambda'$, when working on the N dimensional subspace, we have 
\begin{eqnarray}
\mathcal{H}_i=\left ( \begin{array}{cccccc}
    E_1 & \lambda' & 0 & ...& 0& 0\\
     \lambda& E_2 &\lambda' & ...& 0& 0\\
     0& \lambda &E_3 & ...& 0& 0\\
     \vdots &  \vdots & \vdots &  \ddots&  \vdots&  \vdots\\
     0& 0 &0 & ...& E_{N-1}& \lambda'\\
     0& 0 &0 & ...& \lambda& E_N\\
\end{array}\right)=H+D_N
\label{app_H1}
\end{eqnarray}
In the interaction picture, we have
\begin{eqnarray}
    &&D_{NI}(t)=e^{-iHt}D_Ne^{iHt}\nonumber\\
    &=&\left ( \begin{array}{cccc}
    0 & \lambda'e^{i(E_2-E_1)t} & 0 & ...\\
     \lambda e^{-i(E_2-E_1)t}& 0 &\lambda'e^{i(E_3-E_2)t} & ...\\
     0& \lambda e^{-i(E_3-E_2)t} &0 & ...\\
     \vdots &  \vdots & \vdots &  \ddots  
\end{array}\right)
\end{eqnarray}
For $\lambda\ll \Delta E$, it is legitimate to keep $\rho(t)$ up to the first order
\begin{eqnarray}
    \rho(t)&\approx& \frac{1}{N}(I_N-i\int_0^t D_{NI}(t')dt'+i\int_0^t D_{NI}^\dagger(t')dt')
    \label{app_rho_t1}\nonumber\\
    &=&\frac{I_N}{N}+R(t)
\end{eqnarray}
with $[R(t)]_{i,i-1}=\frac{\lambda-\lambda'}{\Delta E}(e^{-i\Delta Et}-1)$ and $[R(t)]_{i,i+1}=\frac{\lambda-\lambda'}{\Delta E}(e^{i\Delta Et}-1)$. Therefore, the deviation of $L(t)$ to 1 is of order $\Delta \lambda/\Delta E$, which can be omitted according to our assumption $\Delta E\gg\lambda, \lambda'$.

\section{Appendix C: Kitaev chain model and its error}

For the Kitaev model
\begin{eqnarray}
H=\sum_j^{N-1} [&-&t(a_j^\dagger a_{j+1}+a_{j+1}^\dagger a_j)-\mu(a_j^\dagger a_j-1/2)\nonumber\\
&+&\Delta a_j a_{j+1}
+\Delta^*a_{j+1}^\dagger a_j^\dagger]
\label{App_Kitaev}
\end{eqnarray}
its bulk spectrum is given by\cite{Yu_Kitaev2001-ys}
\begin{eqnarray}
E(k)=\pm \sqrt{(2t\cos k+\mu)^2+4|\Delta|^2\sin^2 k}
\end{eqnarray}
It predicts the phase boundary at $|\mu|=|2t|$. Expressing the electron operators  in terms of Majorana operators $a_j=\frac{1}{2}(c_{2j-1}+ic_{2j})$, the Hamiltonian Eq.~(\ref{App_Kitaev}) can be written as
\begin{eqnarray}
H=\frac{i}{2}\sum_j [&-&\mu c_{2j-1}c_{2j}+(\Delta+t)c_{2j}c_{2j+1}\nonumber\\
&+&(\Delta-t)c_{2j-1}c_{2j+2}]
\label{Kitaev_M}
\end{eqnarray}
When $2t>|\mu|$, it is in the topological phase with two Majorana zero modes given by\cite{Yu_Kitaev2001-ys}
\begin{eqnarray}
\gamma&=&\sum_j(\alpha_+x_+^j+\alpha_-x_-^j)c_{2j-1}\nonumber\\
\gamma'&=&\sum_j(\alpha'_+x_+^{-j}+\alpha'_-x_-^{-j})c_{2j}
\end{eqnarray}
where
\begin{eqnarray}
x_{\pm}=\frac{-\mu\pm\sqrt{\mu^2-4t^2+4\Delta^2}}{2(t+\Delta)}
\label{x_pm}
\end{eqnarray}
The two degenerate ground states $\ket{\psi_0}$ and $\ket{\psi_1}$ satisfy 
\begin{eqnarray}
i\gamma\gamma'\ket{\psi_0}=\ket{\psi_0},\quad i\gamma\gamma'\ket{\psi_1}=-\ket{\psi_1}
\end{eqnarray}
Therefore, $P_\gamma=i\gamma\gamma'$ is the fermionic parity operator for the ground state subspace.

For the errors in Fig.~\ref{fig1}(b) of the main text, they can be traced back to our choice of an approximate $J$ for specific parameters. The relative error of this approximation is of order $|x_\pm|$ in Eq.~(\ref{x_pm}). The shrink of the topological phase region predicted by $\bar{L}$ when $\Delta$ becomes larger in Fig.~\ref{fig1}(b) is due to the fact that $|x_\pm|$ increase with $\Delta$ and thus introduces more error into $H_\delta$. This corresponds to the situation that $\lambda'$ becomes too large in Eq.~(6) therefore $L(t)$ no longer predicts the right phase. 

\section{Appendix D: Obtaining the tight binding Hamiltonian for the nanowire model and phase diagrams for randomly chosen perturbations}

For a one-dimensional semiconductor wire with strong Rashba spin orbit interaction placed in proximity to a superconductor, its Hamiltonian is given by\cite{Lutchyn2010-xj,Oreg2010-qv,Sau2010-wi},
\begin{eqnarray}
H_0=&\int& dk \psi_\sigma^\dagger(k)(\frac{k^2}{2m^*}-\mu +2\alpha k\sigma_z\nonumber\\
&+&V_x\sigma_x)_{\sigma\sigma'}\psi_{\sigma'}(k)
\label{H_semi}
\end{eqnarray}
where $m^*$, $\mu$, $\alpha$, $V_x$ are the effective mass, chemical potential, strength of spin-orbit interaction and Zeeman splitting, respectively. And 
\begin{eqnarray}
H_{SC}=\int dk(\Delta \psi_\uparrow(k)\psi_\downarrow(-k)+H.c.)
\end{eqnarray}
where $\Delta$ is the superconducting pairing potential assumed to be real. 
Transforming a continuous model to a lattice model can be achieved by the replacement\cite{Shen2012-yd}: $k\rightarrow\frac{1}{a} \sin ka$ and $k^2\rightarrow\frac{2}{a^2}(1- \cos ka)$. Regrouping the coefficients, the tight binding Hamiltonian reads
\begin{eqnarray}
H&=&\sum_{k,\sigma,\sigma'}a_{k\sigma}^\dagger(2t-2t\cos k-\mu+2\alpha \sin k\sigma_z\nonumber\\
&+&V\sigma_x)_{\sigma\sigma'}a_{k\sigma'}+
\sum_k \Delta a_{k\uparrow}a_{-k\downarrow}+\Delta a_{-k\downarrow}^\dagger a_{k\uparrow}^\dagger
\label{Hk_spin}
\end{eqnarray}
Transformed backed to the coordinate space, we obtain the desired lattice Hamiltonian
\begin{eqnarray}
H&=&\sum_{i,\sigma}\frac{1}{2}[(2t-\mu)a_{i\sigma}^\dagger a_{i\sigma}-(2t-2i\alpha\sigma)a_{i+1\sigma}^\dagger a_{i\sigma}\nonumber\\
&+&V a_{i\uparrow}^\dagger a_{i\downarrow}+\Delta a_{i\uparrow} a_{i\downarrow}]+H.c.
\label{Hi_spin}
\end{eqnarray}
with spectrum
\begin{eqnarray}
E(k)&=&\pm(\epsilon_k^2+4\alpha^2\sin^2k+\Delta^2+V^2\nonumber\\
&\pm& 2\sqrt{\Delta^2V^2+V^2\epsilon_k^2+4\alpha^2\epsilon_k^2\sin^2k})^\frac{1}{2}
\end{eqnarray}
which gives the same phase boundary $V^2=\Delta^2+\mu^2$ as the continuous model.

Now let's consider what will happen if we probe the system using randomly chosen non-Hermitian perturbations. We consider the following non-Hermitian perturbations that are only a part of the ideal $J$ in the main text,
\begin{eqnarray}
H_{\delta1}&=&ie^{i\phi}a_{1\uparrow}^\dagger-ie^{-i\phi}a_{1\uparrow}-ie^{i\phi}a_{N\downarrow}^\dagger-ie^{-i\phi}a_{N\downarrow}\nonumber\\
H_{\delta2}&=&ie^{i\phi}a_{1\uparrow}^\dagger-ie^{-i\phi}a_{1\uparrow}-e^{-i\phi}a_{N\uparrow}^\dagger+e^{i\phi}a_{N\uparrow}\nonumber\\
H_{\delta3}&=&ie^{i\phi}a_{1\uparrow}^\dagger-ie^{-i\phi}a_{1\uparrow}+e^{-i\phi}a_{1\downarrow}^\dagger+e^{i\phi}a_{1\downarrow}\nonumber\\
H_{\delta4}&=&ie^{-i\phi}a_{N\uparrow}^\dagger-ie^{i\phi}a_{N\uparrow}-e^{i\phi}a_{N\downarrow}^\dagger-e^{-i\phi}a_{N\downarrow}
\end{eqnarray}
where $H_{\delta1}$ contains only spin up components on site 1 and spin down components on site N, $H_{\delta2}$ contains spin up components on site 1 and N, $H_{\delta3}$ and $H_{\delta4}$ contains only components on one edge. Using the same parameters as in Fig.~\ref{fig2}(b), we calculate $\bar{L}$ and obtain the results in Fig.~\ref{fig4}. We find that for $H_{\delta1}$ and $H_{\delta2}$ that contain zero mode components on both edges, the phases are still discernible from the $\bar{L}$, while for $H_{\delta3}$ and $H_{\delta4}$ that contain only components on one edge, the $\bar{L}$ take nearly the same value in both phases thus the method breaks down. Fig.~\ref{fig4} tells us that if we are unable to find out the exact Majorana zero modes, we can still use this dynamical method to differentiate the topological trivial and non-trivial phases by choosing a non-Hermitian perturbation that contains contributions from both edges. This robustness therefore renders some flexibility to the experimental realizations.

\begin{figure}[!t]
    \centering
    \includegraphics[width=7in]{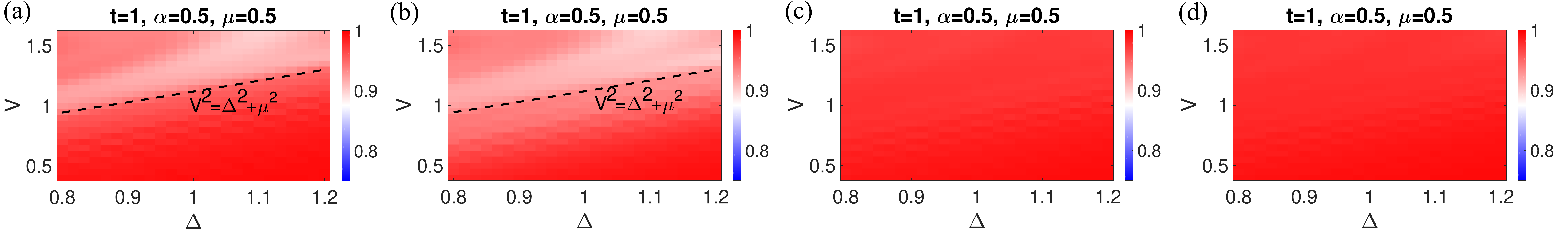}
    \caption{(a)(b)(c)(d) correspond to the phase diagram obtained using $H_{\delta1}$, $H_{\delta2}$, $H_{\delta3}$, $H_{\delta4}$ respectively. Parameters: $\phi=0.36$, $t=1$, $\mu=0.5$, $\alpha=0.5$, $\beta=5$, $N=4$, $\lambda=0.05$. For $H_{\delta1}$ and $H_{\delta2}$ which contain components from both edges, the phases are still discernible from the $\bar{L}$. For $H_{\delta3}$ and $H_{\delta4}$ which contain components from a single edge, the phases are undiscernible from the $\bar{L}$. Parameters are the same as Fig.~(2b) in the main text. }
    \label{fig4}
\end{figure}



\end{widetext}


\end{document}